\documentclass{aa}
\usepackage{graphicx}
\usepackage{txfonts}

\begin{document}

\title{Ionization structure in the winds of B[e] supergiants}
\subtitle{II. Influence of rotation on the formation of equatorial hydrogen neutral zones}

\author{Michaela Kraus\inst{1,2}}

\offprints{M. Kraus, \\ \email{kraus@sunstel.asu.cas.cz}}

\institute{Astronomick\'y \'ustav, Akademie v\v{e}d \v{C}esk\'e republiky, Fri\v{c}ova 298, 251~65 Ond\v{r}ejov, Czech Republic\\
           \email{kraus@sunstel.asu.cas.cz}
\and
Sterrekundig Instituut, Universiteit Utrecht, Princetonplein 5, 3584 CC Utrecht, The Netherlands\\
      }

\date{Received / accepted}

% abstract : {Context}{Aims}{Methods}{Results}{Conclusions}

\abstract
{B[e] supergiants are known to have non-spherical winds, and the existence
of disks that are neutral in hydrogen close to their stellar surface
has been postulated. A suitable mechanism to produce non-spherical
winds seems to be rapid rotation, and at least for three B[e] supergiants in 
the Magellanic Clouds rotation velocities at a substantial fraction of their
critical velocity have been found.}
{We want to find suitable recombination distances in the equatorial
plane of rapidly rotating stars that explain the observed 
huge amounts of neutral material in the vicinity of B[e]
supergiants.}
{We perform ionization structure calculations in the equatorial plane around
rapidly rotating luminous supergiants. The restriction to the equatorial plane
allows us to treat the ionization balance equations 1-dimensionally, while
the stellar radiation field is calculated 2-dimensionally, taking into account
the latitudinal variation of the stellar surface parameters. 
The stellar parameters used correspond to those known for B[e] supergiants. 
The assumptions made in the 
computations all have in common that the total number of available
ionizing photons 
at any location within the equatorial plane is overestimated, resulting
in upper limits for the recombination distances.}
{We find that despite the drop in equatorial surface density of rapidly rotating
stars (neglecting effects like bi-stability and/or wind compression), 
hydrogen and helium recombine at or close to the 
stellar surface, for mass loss rates $\dot{M} \ga 5\times 
10^{-5}$M$_{\odot}$yr$^{-1}$ and rotation speeds in excess of 
$v_{\rm rot, eq}/v_{\rm crit} \simeq 0.8$.}
{}


\maketitle

\section{Introduction}

In a series of papers Maeder \& Meynet discussed the importance and the 
influence of rapid rotation on the evolution of stars, the chemical yields
and non-spherical mass and angular momentum loss (see e.g Maeder, 
\cite{Maeder99}; Meynet \& Maeder, \cite{MeMa00}; Maeder \& Meynet, 
\cite{MaMe00}). The influence of rotation on so many stellar parameters
results also in the shaping of the wind and the nebula.  There 
has been the suggestion that the appearance of non-spherical winds around 
some massive and luminous stars might be caused by rotation (Maeder 
\cite{Maeder02}; Maeder \& Desjacques \cite{MaDes}). Stars for which this might 
be appropriate are the Luminous 
Blue Variables and the B[e] supergiants. 
%Especially this latter group of stars 
%has gained increasing attraction during the last $\sim 20$ years. 

Zickgraf et al. (\cite{Zickgraf85}) suggested that the hybrid character of 
the optical spectra of B[e] supergiants is due to a non-spherical (two-component) wind. The 
strong infrared excess indicates the presence of a huge amount of hot 
circumstellar dust, and polarimetric observations, e.g. by Magalh\~aes 
(\cite{Magalhaes}), Magalh\~aes et al. (\cite{Magalhaesetal}), and Melgarejo et 
al. (\cite{Melgarejo}), confirmed the non-spherical geometry of the 
circumstellar material around B[e] supergiants. The existence of a 
geometrically thick circumstellar disk responsible for the polarized emission 
and the location of the hot dust seems nowadays to be well established 
(for an overview see e.g. Kraus \& Miroshnichenko \cite{KM06}). The formation 
mechanism of these disks is, however, still rather unclear. There exist two 
promising approaches: (1) the bi-stability mechanism introduced by Lamers \& 
Pauldrach (\cite{LP}) and further investigated, especially with respect to the 
influence of rotation on the formation of B[e] supergiant stars' disks, by 
Pelupessy et al. (\cite{Pelupessy}), and (2) the wind-compressed disk, 
introduced by Bjorkman \& Cassinelli (\cite{BC}). Both models however have
difficulties in explaining {\sl all} observed quantities of the 
B[e] supergiants' disks in a self-consistent way. In addition, there is still 
no consensus about the nature of these disks, whether they can be described by 
a high density equatorial outflowing wind or by a Keplerian viscous disk (see 
e.g. Porter \cite{Porter}; Kraus \& Miroshnichenko \cite{KM06}).

Recently, these disks have been suggested to be neutral in hydrogen in 
the vicinity of the stellar surface. Tracers for hydrogen neutral material are 
e.g. the strong [O{\sc i}] emission lines arising in the optical spectra of 
B[e] supergiants. Modeling of their line luminosities revealed that, in
order to keep the
mass loss rate of the star at a reliable value, these lines must originate 
within a few stellar radii from the surface (Kraus \& Borges Fernandes 
\cite{KrausBorges}; Kraus et al.  \cite{Krausetal06}). In addition, several
B[e] supergiants are found to show band-head emission from hot (3000 - 5000\,K)
CO gas (McGregor et al. \cite{McGregor}). Follow-up studies of 
high-resolution spectra for at least one of them led to the conclusion that
this hot CO gas is located at about 2-3 AU from the hot stellar surface
(Kraus \cite{PhD}; Kraus et al. \cite{Kraus00}). 
%And also around the Luminous Blue Variable star AG Car, rather cool CO 
%emission has been detected (Nota et al. \cite{Nota}).
The existence of neutral material close to these luminous 
B[e] supergiants is surprising and needs to be investigated in detail. The goal 
of our study is therefore to find scenarios that allow neutral material to 
exist close to the surface of these stars.

In a first attempt, Kraus \& Lamers (\cite{KrausLamers}, hereafter Paper\,1) 
calculated the ionization structure of B[e] supergiants, assuming a 
latitude--dependent mass flux that increases from pole to equator. With such a model they
could show that, even with moderate total mass loss rates, hydrogen recombines
in the equatorial direction close to the star leading to a hydrogen neutral
disk-like structure. 
Here, we investigate the influence of rotation 
on the stellar parameters and consequently on the ionization structure in the
winds and disks of B[e] supergiants. 

Rotation causes a flattening of the stellar surface and therefore a reduction
of the local net gravity in the equatorial region. The decrease in gravity from
pole to equator equally results in a decrease of the stellar flux which
is proportional to the local net gravity. Hence, the effective temperature
also decreases from pole to equator, known as gravity darkening 
(or polar brightening, von Zeipel \cite{vonZeipel}).
The latitude dependence of the gravity and effective temperature has also
impacts on the stellar wind parameters (see e.g. Lamers \& Cassinelli 
\cite{LamersCassinelli}): The escape velocity following from the balance 
between gravitational and centripetal forces becomes latitude--dependent, 
decreasing from pole to equator. The same holds for the terminal wind velocity
which is (for line-driven winds) proportional to the escape velocity.
Even the mass flux from the star tends to decrease from pole to equator
if gravity darkening is taken into account in the CAK theory as shown by 
Owocki et al. (\cite{Owocki98}).
More important for the ionization structure calculations is the density in
the wind, and we will show in Sect.\,\ref{basics} that the density at 
any given distance also {\it decreases} from pole to equator. A rotating star 
will therefore have a less dense wind in the equatorial region, unless
special effects such as bi-stability or wind compression play a role.

Both important parameters in the ionization balance equations, i.e. the
effective surface temperature {\it and} the surface density, decrease from pole
to equator. While the decrease in surface temperature tends to decrease the
number of available photons suitable to ionize H and He, the decrease in
surface density reduces the optical depth along the line of sight
from a point in the wind to the star. Both effects are therefore counteracting
with respect to the location where recombination takes place. While a reduction
of ionizing photons will shift the recombination distance towards the star,
the reduction in optical depth along the direction to the star will shift it
further outwards. The outcome of the ionization balance equations is therefore
unpredictible and very sensitive to the chosen input parameters. We thus
investigate the ionization structure in the wind of a rotating star in
more detail.

The paper is structured as follows: In Sect.\,\ref{basics} we provide the 
equations that describe the surface distribution of the effective temperature,
mass flux, escape velocity (and hence terminal wind velocity), and hydrogen
density for a rigidly rotating star. The ionization structure calculations
restricted to the equatorial plane of the systems are performed in 
Sect.\,\ref{wind} where also the results for the recombination distances
of helium and hydrogen are shown. The influence of the assumptions on these 
results and the applicability of the models to B[e] supergiants are discussed 
in Sects.\,\ref{discuss} and \ref{applic}, respectively, and the conclusions 
are given in Sect.\,\ref{conclusions}.

%%%%%%%%%%%%%%%%%%%%%%%%%%%%%%%%%%%%%%%%%%%%%%%%%%%%%%%%%%%%%%%%%%%%%%%%%
                                                                                
\section{The surface and wind structure of rigidly rotating stars}\label{basics}

In this paper, we restrict our investigations to rigid rotation only, and 
we neglect any influences due to bi-stability and wind-compression.

\subsection{The shape of the stellar surface}
                                                                                
%To calculate the latitude dependence of the stellar radius, $R(\theta)$, we 
%start with the Roche model for the equipotential surfaces. 
The potential $\Phi$ of a rotating star is given by the sum of  
gravitational and centrifugal potential. The latitude dependence 
of the stellar radius, $R(\theta)$, i.e. the shape of the star, is determined
by the equipotential surfaces, $\Phi(R(\theta),\theta,\phi)$, for which
it is assumed that all the mass is concentrated in the core. These
equipotential surfaces are given by
\begin{equation}
\Phi(R(\theta),\theta,\phi) = -\frac{GM_{\rm eff}}{R(\theta)} - \frac{1}{2}R^{2}(\theta)\Omega^{2}\sin^{2}(\theta)
\label{Phi_tot}
\end{equation}
where $\theta$ is the co-latitude with $\theta = 0$ at the pole, $M_{\rm eff}$ 
is the effective stellar mass, i.e. the stellar mass reduced by the 
effects of radiation pressure due to electron scattering, and $\Omega$ is the 
angular velocity. With the
definitions of $x(\theta)=R(\theta)/R_{\rm eq}$, $v_{\rm crit} = \sqrt{(GM_{\rm 
eff})/R_{\rm eq}}$, and $\omega = v_{\rm rot, eq}/v_{\rm crit}$, the 
latitude--dependent stellar radius is found from Eq.\,(\ref{Phi_tot}) which 
results in the following cubic function
\begin{equation}
x^{3} - \frac{2+\omega^{2}}{\omega^{2}\sin^{2}\theta}\,x + \frac{2}
{\omega^{2}\sin^{2}\theta} = 0
\label{cubic}
\end{equation}
with the solution
\begin{equation}\label{R_pol_R_eq}
x(\theta = 0) = \frac{R_{\rm pole}}{R_{\rm eq}} = \left(1 + \frac{1}{2}\omega^{2}
\right)^{-1}
\end{equation}
and
\begin{equation}\label{x_theta}
x(\theta\neq 0) = 2\frac{\sqrt{2 + \omega^{2}}}{\sqrt{3} \omega\sin\theta}\,\sin
\left\{\frac{1}{3}\arcsin\left(\frac{3\sqrt{3} \omega\sin\theta}{(2+
\omega^{2})^{3/2}}\right)\right\}\,.
\end{equation}
Equations (\ref{R_pol_R_eq}) and (\ref{x_theta}) describe the stellar radius 
at all latitudes for a star rotating rigidly with a specific value of $\omega$.

\subsection{The latitude--dependent surface temperature}

Rotation not only influences the radius of the star but also results in a
latitude--dependent surface temperature distribution, because the stellar
flux, $F$, is proportional to the local effective gravity, $g_{\rm 
eff}$, which is calculated from
\begin{equation}
g_{\rm eff} = -\nabla \Phi \sim \frac{1}{R^{2}(\theta)}\,\left(1-\frac{R^{3}
(\theta)}{R_{\rm eq}^{3}}\omega^{2}\sin^{2}\theta\right)\,.
\end{equation}
Since $F(\theta) = \sigma T_{\rm eff}^{4}(\theta) \sim g_{\rm eff}$\,,
the surface temperature $T_{\rm eff}(\theta)$ behaves as
\begin{equation}\label{T_prop}
T_{\rm eff}^{4}(\theta)\sim\frac{1}{R^{2}(\theta)}\,\left(1-\frac{R^{3}(\theta)}{R_{\rm eq}^{3}}\omega^{2}\sin^{2}\theta\right)
\end{equation}
or, if we express the latitude--dependent effective temperature in terms
of the polar temperature, $T_{\rm eff}({\rm pole})$, replace $R(\theta)$
by $x(\theta)$, and make use of relation (\ref{R_pol_R_eq}),
\begin{eqnarray}
T_{\rm eff}^{4}(\theta) & = & T_{\rm eff}^{4}({\rm pole})\frac{R^{2}({\rm
   pole})}{R^{2}(\theta)}\,\left(1-\frac{R^{3}(\theta)}{R_{\rm eq}^{3}}
   \omega^{2}\sin^{2}\theta\right) \nonumber\\
 & = & T_{\rm eff}^{4}({\rm pole})\frac{1}{x^{2}(\theta)}\frac{(1-x^{3}(\theta)
\omega^{2}\sin^{2}\theta)}{(1+\frac{1}{2}\omega^{2})^{2}}\,.
\label{T_theta}
\end{eqnarray}

\begin{figure}[t!]
\resizebox{\hsize}{!}{\includegraphics{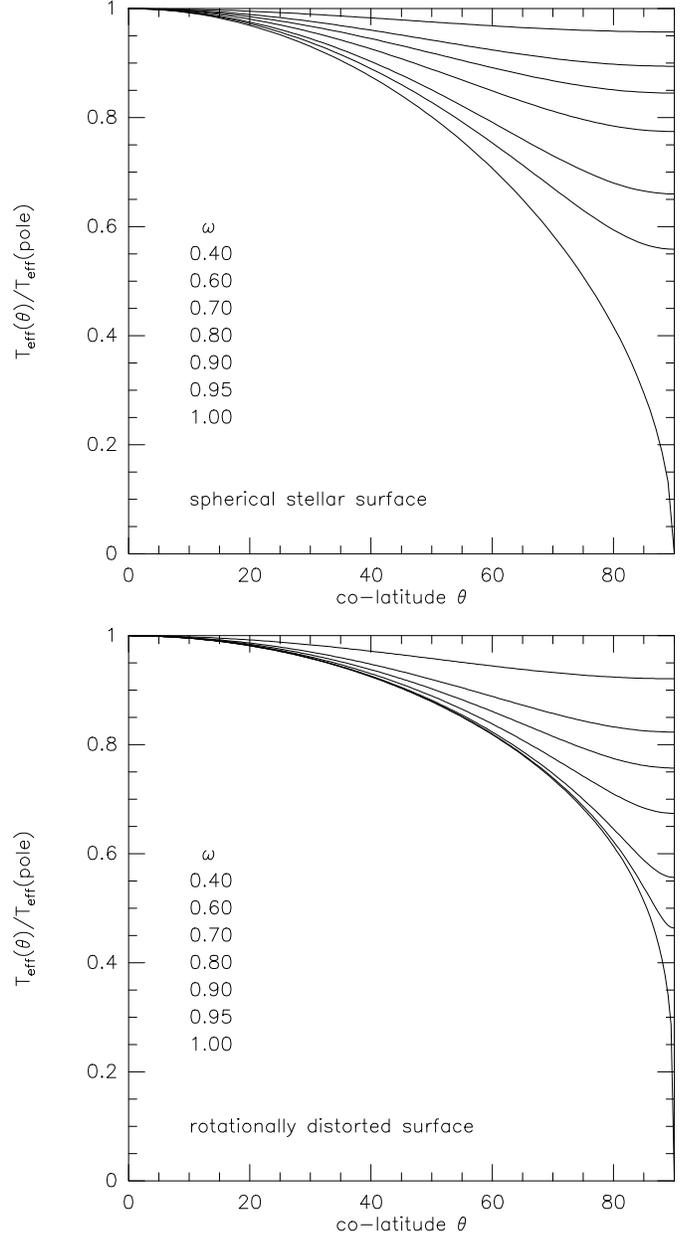}}
\caption{Effective temperature distribution on the surface of a rotating
star. The different curves, which are normalized to the polar temperature, are 
for different rotational velocities indicated by $\omega$
from low values (upper curves) to high values (lower curves).
In the top panel the 
rotational distortion of the stellar surface has been neglected.} 
\label{T_eff}
\end{figure}

For the purpose of our paper it is important to treat the surface effective
temperature (as well as all other following parameters) properly, which means
that we have to take into account the rotationally distorted stellar surface.
How different the results can be when accounting only for gravity darkening but 
neglecting the real shape of the star is shown in Fig.\,\ref{T_eff}. There we 
compare the surface temperature distribution calculated from 
Eq.\,(\ref{T_theta}) which accounts for the distorted surface with the
one resulting from a rotating star but under the assumption of an unperturbed, 
spherical surface (i.e. $R(\theta) = R_{\rm eq} = R$). In this latter case, 
$T_{\rm eff}(\theta)$ resulting from Eq.\,(\ref{T_prop}) is simply given by
(see e.g. Lamers \& Cassinelli \cite{LamersCassinelli}; Lamers\footnote{Please 
note the typo in Lamers' Eq.\,(5) where it should be $\cos^{2}(\theta)$
instead of $\cos(\theta)$, and that in his paper $\theta$ is measured 
from the equator.} \cite{Lamers04})
\begin{equation}
T_{\rm eff}^{4}(\theta) = T_{\rm eff}^{4}({\rm pole})\left(1 - \omega^{2}
\sin^{2}\theta \right)\,.
\end{equation}
This equation also describes globally the influence of rigid rotation, i.e. the
drop in temperature from pole to equator. However, the absolute value of the
effective temperature at any location on the stellar surface is different,
as is obvious from the comparison in Fig.\,\ref{T_eff}: 
\begin{itemize}
\item For $\omega \la 0.8$ the rotationally distorted surface is cooler at 
all latitudes.
\item For $\omega \ga 0.8$  
the effective temperature of the rotationally distorted surfaces 
is higher for small to intermediate latitudes, but becomes (much) lower in 
the equatorial regions, compared to the corresponding spherical surfaces.
\end{itemize}
These severe differences in surface temperature distribution have non-negligible
effects on the stellar radiation field at any point in the wind. A proper 
treatment of the stellar parameters by accounting for the rotationally 
distorted surface is therefore an important ingredient in our ionization 
balance calculations.

\subsection{The latitude--dependent mass flux}
                                                                                
For the mass flux, $F_{\rm m}$ we follow the description of Owocki et al. 
(\cite{Owocki98}) given by their Eq.\,(2)
\begin{equation}
\frac{F_{\rm m}(\theta)}{F_{\rm m}({\rm pole})} = \left[ \frac{F(\theta)}
{F({\rm pole})}\right]^{\frac{1}{\alpha}}\,\left[ \frac{g_{\rm eff}(\theta)}
{g_{\rm eff}({\rm pole})}\right]^{1-\frac{1}{\alpha}}\,.
\end{equation}
This equation describes the latitude--dependent mass flux according to CAK 
theory (Castor et al.\,\cite{CAK}). Neglecting bi-stability effects, which 
means that the force multiplier $\alpha$ is constant all over the surface, 
and introducing gravitational darkening according to the von Zeipel theorem 
(i.e. $F(\theta) \sim g_{\rm eff}(\theta)$) results in
\begin{equation}
\frac{F_{\rm m}(\theta)}{F_{\rm m}({\rm pole})} = \frac{g_{\rm eff}(\theta)}
{g_{\rm eff}({\rm pole})} = \frac{1}{x^{2}(\theta)}\,\frac{(1-x^{3}(\theta)
\omega^{2}\sin^{2}\theta)}{(1+\frac{1}{2}\omega^{2})^{2}}\,.
\label{Fm_theta}
\end{equation}
The surface distribution of the mass flux for different values of $\omega$ is 
shown in the top panel of Fig.\,\ref{surf}.

\subsection{The latitude--dependent terminal velocity}
                                                                            
\begin{figure}[t!]
\resizebox{\hsize}{!}{\includegraphics{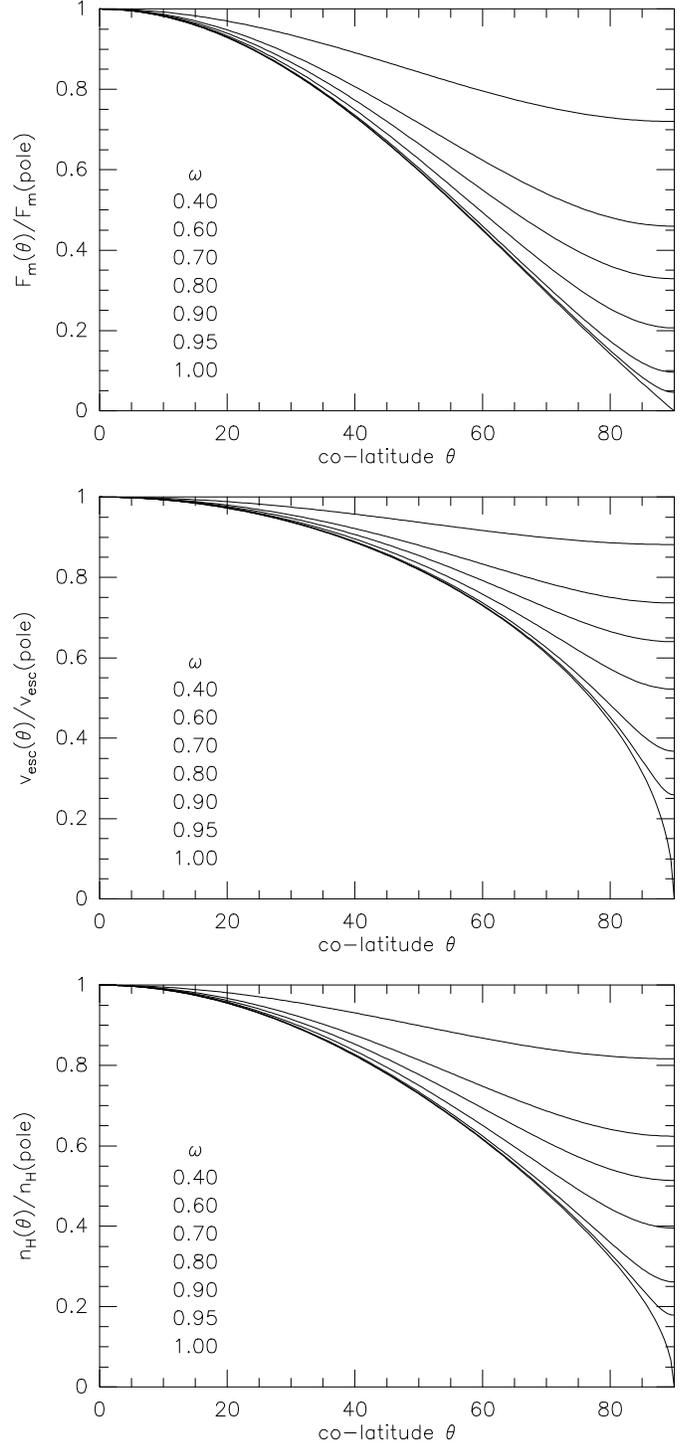}}
\caption{Distribution of the mass flux (top panel), the escape velocity (mid panel), and the hydrogen density (bottom panel) on the surface of a rotating
star. The different curves in each plot, which are normalized to the 
corresponding polar value of the parameters, are for different rotational
velocities indicated by $\omega$. All parameters drop from pole to equator. This
effect becomes stronger with increasing values of $\omega$.}
\label{surf}
\end{figure}

The escape velocity of a rotating star follows from balancing gravitational 
and centripetal forces on the stellar surface which means that the effective 
gravity $g_{\rm eff}(\theta)$ must equal $v_{\rm esc}^{2}(\theta)/R(\theta)$.
Since the terminal wind velocity, $v_{\infty}$, is, according to line-driven 
wind theories (see e.g. Lamers \& Cassinelli \cite{LamersCassinelli}, Chapter 
8), proportional to the escape velocity, $v_{\rm esc}$, we find the following
relation for the latitude dependence of the terminal velocity
\begin{equation}
v_{\infty}(\theta) = v_{\infty}({\rm pole}) 
\frac{\left( 1 - x^{3}(\theta) \omega^{2}\sin^{2}\theta \right)^{1/2}}
{\left(x(\theta) (1+\frac{1}{2} \omega^{2} ) \right)^{1/2}}\,.
\label{v_inf_theta}
\end{equation}
The latitude dependence of the terminal velocity for different values of 
$\omega$ is shown in the middle panel of Fig.\,\ref{surf}.

\subsection{The latitude--dependent wind density distribution}

It is known that the mass flux and the terminal velocity 
decrease from pole to equator, even if the rotational distortion of the
stellar surface is neglected in their derivation (see Lamers \& Cassinelli
\cite{LamersCassinelli}). For the ionization structure calculations, however, 
we need to know the density distribution in the wind. 

In a non-rotating, spherically symmetric stationary wind, the density at any 
location $r$ in the wind is related to the mass loss rate, $\dot{M}$, of the 
star and the wind velocity, $v(r)$, via the equation of mass continuity 
\begin{equation}
n_{\rm H}(r) = \frac{\dot{M}}{4\pi\mu m_{\rm H}r^{2}v(r)} = \frac{F_{\rm m}}
{\mu m_{\rm H} v(r)}\,\frac{R_{*}^{2}}{r^{2}}
\label{masscont}
\end{equation}
where $\mu$ is the mean molecular weight and $n_{\rm H}$ denotes the 
particle density of hydrogen given in cm$^{-3}$.  
From the right-hand side of this equation it follows immediately that in a 
non-spherically symmetric wind, the radial density distribution at any latitude 
can be written in the form
\begin{equation}
n_{\rm H}(\theta,r) = \frac{F_{\rm m}(\theta)}{\mu m_{\rm H} v(\theta,r)}\,
\frac{R^{2}(\theta)}{r^{2}}\,.
\label{dens}
\end{equation}
For our further calculations, we assume that the wind velocity is constant
in radial direction, i.e. $v(\theta,r) = v(\theta,R) = v(\theta)$. Therefore
we can re-write Eq.\,(\ref{dens}) in the form
\begin{equation}
n_{\rm H}(\theta,r) = n_{\rm H}(\theta,R(\theta)\,)\,\frac{R^{2}
(\theta)}{r^{2}}
\end{equation}
where 
\begin{equation}
n_{\rm H}(\theta,R(\theta)\,) = \frac{F_{\rm m}(\theta)}{\mu m_{\rm H}
v(\theta)}
\end{equation}
defines the density distribution along the stellar surface.

With the additional simplification of $v(\theta) = v_{\infty}(\theta)$,
we can express the surface density distribution,
$n_{\rm H}(\theta,R(\theta)\,)$, by using the relations for the mass
flux given by Eq.\,(\ref{Fm_theta}) and for the terminal velocity given by 
Eq.\,(\ref{v_inf_theta})  
\begin{eqnarray}
n_{\rm H}(\theta,R(\theta)\,) & = &\frac{F_{\rm m}({\rm pole})}{\mu m_{\rm H}v_{\infty}({\rm
pole})}\,\frac{\left( 1 - x^{3}(\theta) \omega^{2}\sin^{2}\theta \right)^{1/2}}
{\left(x(\theta) (1+\frac{1}{2}\omega^{2} ) \right)^{3/2}} \\
 & = & n_{\rm H}({\rm pole})\,\frac{\left( 1 - x^{3}(\theta) \omega^{2}\sin^{2}\theta \right)^{1/2}}
{\left(x(\theta) (1+\frac{1}{2}\omega^{2} ) \right)^{3/2}}\,.
\end{eqnarray}
This surface density distribution, as the result of the ratio of mass 
flux to terminal velocity, is plotted for different values of $\omega$ in
the lower panel of Fig.\,\ref{surf}. It also decreases from pole to equator.
This means that a rigidly rotating star will have a {\sl less dense wind} in 
the equatorial region, unless bi-stability and wind compression play a role.

\section{Ionization structure calculations}\label{wind}

Since we are searching for the existence of a hydrogen neutral equatorial 
region, we restrict our calculations to the equatorial plane only. This 
leads to the simplification of a symmetrical stellar radiation field with
respect to the equatorial plane. Therefore, it is sufficient to solve the 
ionization balance equations along one radial direction, which we will call
the $y$-axis. 

As in Paper\,1, our model wind consists of hydrogen and helium, only. This 
means that we have to solve two coupled ionization balance equations that
are treated in the on-the-spot (OTS) approximation. This approximation states 
that every photon generated via 
recombination and able to ionize hydrogen or helium will be absorbed 
immediately in the close vicinity of its generation location. The ionization 
balance equations for this case are given in Sect.\,4. of Paper\,1.
The recombination distance is found by applying a root-finding routine.
Usually, a few iteration steps are sufficient for an accuracy in distance
better than 1\%.

\subsection{The stellar radiation field}

Besides the OTS approximation, which defines the diffuse radiation field, we 
need
to calculate the stellar radiation field at any point in the equatorial plane,
or, due to the symmetry in our case, at any point along the $y$-axis.
Differently from the treatment in Paper\,1 we no longer use the assumption
that the star is a point source. The stellar parameters of a rotating star 
vary strongly over the stellar surface, especially for increasing stellar 
rotation. We therefore calculate the stellar radiation by integrating the 
latitude--dependent surface flux over the rotationally distorted stellar 
surface. This is done in the following way:
\begin{itemize}
\item We define the stellar input parameters $T_{\rm eff}({\rm pole})$, $g_{\rm
eff}({\rm pole})$, $v_{\infty}({\rm pole})$, $R_{*}({\rm sphere})$, 
$F_{\rm m}({\rm pole})$.
\item We define the rotation velocity, $\omega$.
\item With $\omega$ and $R_{*}({\rm sphere})$ we calculate the shape of
the stellar surface, i.e. $x(\theta)$. We thereby make use of 
Eq.\,(\ref{R_pol_R_eq}) and of the
mass conservation that relates the spherical radius to the equatorial
and polar radii via $R_{*}^{3} = R_{\rm eq}^{2}\,R_{\rm pole}$\,.
\item At each location $r$ along the $y$-axis we determine the
  angular extent of the stellar surface and its shape. 
This defines the size of the stellar surface (or the surface 
segment) from which radiation will arrive at point $r$.
\item Along this stellar surface segment we calculate the distribution 
of $T_{\rm eff}(\theta)$ and $g_{\rm eff}(\theta)$, and the resulting 
radiation temperature $T_{\rm rad}(\theta)$ (see Sect.\,\ref{radtemp}). 
\item The stellar flux as a function of latitude is then approximated by 
$B_{\nu}(T_{\rm rad}(\theta))$.
\item The total stellar radiation field at point $r$ along the $y$-axis follows
from integration of $B_{\nu}(T_{\rm rad}(\theta))$ over the segment of the 
rotationally distorted surface.
\item To account for optical depth effects, we calculate the {\it minimum} optical
depth, which occurs along the $y$-axis (because of the shortest distance
and the lowest density). This optical depth is adopted for all
directions towards the stellar surface.
\end{itemize}

In this calculation of the stellar radiation field at any point $r$ along the 
$y$-axis we make one important approximation, which is the adoption of the
minimum optical depth towards all directions from $r$ to the stellar surface.
This assumption results in an overprediction of the available ionizing photons,
because the stellar radiation from the higher latitudes with higher $T_{\rm 
eff}$ will be absorbed less. The resulting recombination distance will 
therefore be overestimated.

\subsection{The radiation temperature}\label{radtemp}
                                                                                
The stellar radiation temperature is defined in our calculations as
the Planck temperature which describes the part of the stellar spectrum
that delivers the ionizing photons, i.e. the spectrum shortwards
of 912\,\AA.~The determination of the latitude dependence of the radiation
temperature is not straightforward. We therefore briefly explain how we 
calculate it.

We start with Kurucz model atmospheres (Kurucz \cite{Kurucz}) for 
solar metallicity stars and $\log g$ values between 2.0 and 3.5. We do not
investigate higher values of $\log g$ because we are mainly interested in 
giants and supergiants. For each model atmosphere in this $\log g$ range and
for all available effective temperatures we fitted a Planck function 
to the spectrum shortwards of 912\,\AA.~This delivers the radiation 
temperature for the corresponding ($T_{\rm eff}$, $g_{\rm eff}$) combination.
The radiation temperature is therefore a function of these two parameters, i.e.
$T_{\rm rad} = T_{\rm rad}(T_{\rm eff}, g_{\rm eff})$, and we found the 
following useful parametrization:
\begin{equation}
\log T_{\rm rad} = A(\log g)\,\log T_{\rm eff} + B(\log g)
\label{T_rad}
\end{equation}
with the two functions $A(\log g)$ and $B(\log g)$ given by
\begin{eqnarray}
A(\log g) & = & 1.222 - 0.058 \log(\log g - 1.9) \\
B(\log g) & = & 0.235 \log(\log g - 1.89) - 1.13
\end{eqnarray}
in the range of $8\,000 \la T_{\rm eff} \la 30\,000$ K. The error in 
radiation temperature introduced by this fitting procedure is less than 5\% for
the higher values of $T_{\rm eff}$, and less than 3\% for the lower ones.
We thus can compute 
$T_{\rm rad}(\theta)$ for any combination of $T_{\rm eff}$ and $g_{\rm
eff}$ values that will occur along the surface of a rigidly rotating star,
and hence we can calculate the appropriate stellar radiation at all locations
on the stellar surface. 

Since $T_{\rm rad}(\theta)$ is a function of both $T_{\rm eff}$ and $g_{\rm 
eff}$, its latitude dependence will be different from the one of the effective
temperature (see Sect.\,\ref{results}), and will also vary for stars with 
different stellar parameters.

\subsection{Description of the model stars and their winds}

For our calculations we chose stars with a polar effective temperature
of $T_{\rm eff, pole} \simeq 24\,500$\,K and $\log g_{\rm eff, pole} = 3.5$. 
According to Eq.\,(\ref{T_rad}) this combination results in a polar radiation 
temperature of $T_{\rm rad, pole} \simeq 17\,000$\,K.
 
The radius of the non-rotating star is fixed at $R_{*} = 82\,$R$_{\odot}$.
Together with the chosen effective temperature this results in a luminosity of 
the non-rotating star of $L_{*} \simeq 2.2\times 10^{6}$\,L$_{\odot}$, which 
places the star in the B-type supergiant region within the HR diagram. 

\begin{figure}[t!]
\resizebox{\hsize}{!}{\includegraphics{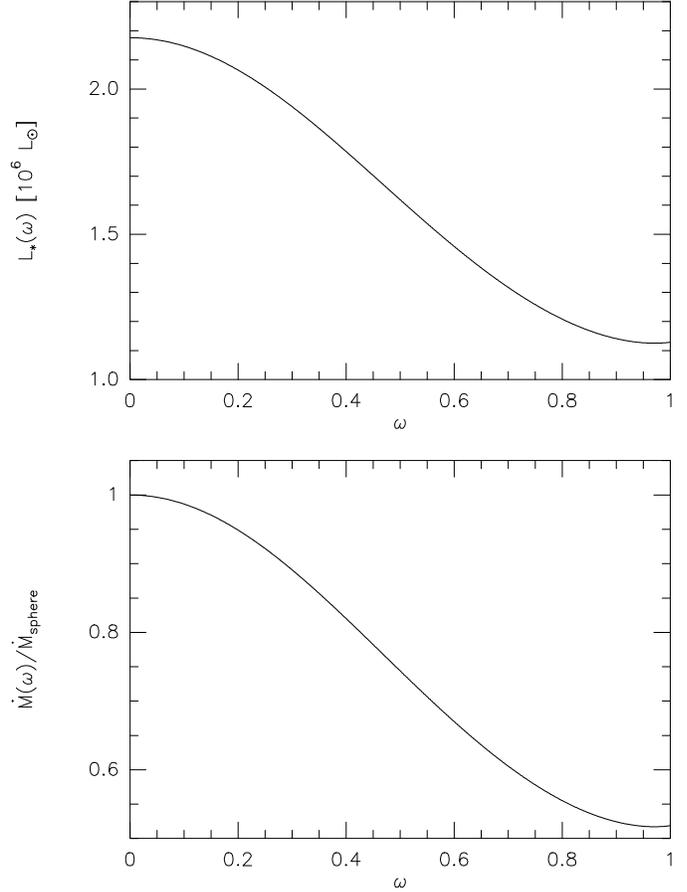}}
\caption{Top panel: Stellar luminosity as a function of $\omega$ for our model 
stars with $T_{\rm eff, pole} = 24\,500$\,K and $R_{*} = 82\,R_{\odot}$.
Bottom panel: Mass loss rate as a function of $\omega$, normalized to the
spherical mass loss rate. 
Both parameters show a decrease with increasing rotation speed. The difference
between the non-rotating and the critically rotating star is about a
factor 2.
}
\label{lum_mdot}
\end{figure}

The polar values of the effective temperature and gravity (and hence radiation 
temperature) are fixed for all our model calculations. Fixing the polar 
effective temperature means that we are {\it not} calculating a star that is 
spinning up. This would result in an increase of polar temperature with 
increasing rotation velocity. Instead, we are calculating stars with the same 
polar effective temperature having different rotation velocities. This means 
that we are dealing with stars of {\it different luminosities}. The total 
luminosity of a star is 
\begin{equation}
L_{*} = \int\limits_{0}^{\pi}\int\limits_{0}^{2\pi}\sigma T_{\rm
eff}^{4}(\theta) \,dS
\end{equation}
where $dS$ is the surface element of an ellipsoid, given by
\begin{equation}
dS = \sqrt{(R_{\rm eq}^{2}\cos^{2}\theta + R_{\rm pole}^{2}\sin^{2}\theta)}~
~R_{\rm eq}\sin\theta\,d\theta~d\phi~.
\end{equation}
The difference in luminosity of rotating stars having the same
polar effective temperature is shown in the upper panel of Fig.\,\ref{lum_mdot} 
where we plotted the distribution of the stellar luminosities as a function of 
$\omega$. With increasing rotation speed, the stellar luminosity drops. The 
difference is largest between the non-rotating and the critically rotating star,
and is about a factor 2.

Similarly, we can calculate the mass loss rates of our model stars.
The mass loss rate follows from
\begin{equation}
\dot{M}_{*} = \int\limits_{0}^{\pi}\int\limits_{0}^{2\pi}F_{\rm m}(\theta)\,dS
\end{equation}
and the ratio of the mass loss rate over the mass loss rate of the spherical 
star is plotted as a function
of $\omega$ in the lower panel of Fig.\,\ref{lum_mdot}.
The mass loss rate shows a difference of about a factor 2 between the
non-rotating and the critically rotating star. The behaviour of both the 
luminosity and the mass loss rate with $\omega$ is identical because
both $T_{\rm eff}^{4}(\theta)$ and $F_{\rm m}(\theta)$ are proportional
to $g_{\rm eff}(\theta)$ (see Eqs.\,(\ref{T_theta}) and (\ref{Fm_theta}) 
respectively). We can therefore conclude that an almost critically 
rotating star will have only half the total luminosity and half the total mass
loss rate of its non-rotating counterpart, having both the same polar
mass flux and effective temperature.  

In our calculations, the polar mass flux, $F_{\rm m, pole}$, is a free 
parameter. Its value is 
varied over a large range to investigate the ionization structure of rotating 
stars and to find the recombination distances (see Sect.\,\ref{results}).

We further use a distant independent wind velocity, which is set to the terminal
velocity, i.e. $v(\theta, r) = v_{\infty}(\theta)$. For the polar wind velocity
we adopt $v_{\infty, \rm pole} = 2000$\,km\,s$^{-1}$. The radial 
electron temperature distribution in the wind is assumed to be constant, and we 
set $T_{\rm e}(\theta,r) = T_{\rm e}(\theta) = 0.8\,T_{\rm eff}(\theta)$. 
The influence of these assumptions and simplifications on the results
are discussed in Sect.\,\ref{discuss}.

\subsection{Recombination in the equatorial plane}\label{results}

\begin{figure}%[h!]
\resizebox{\hsize}{!}{\includegraphics{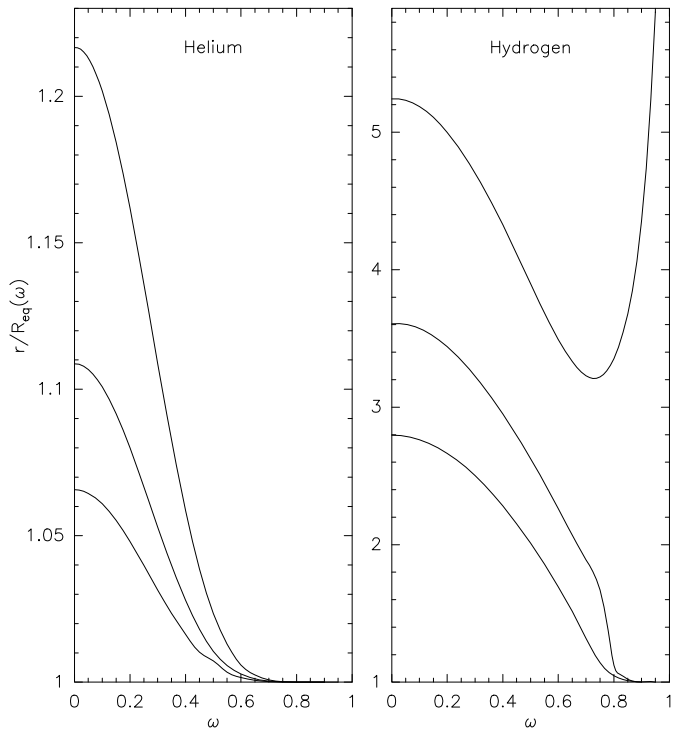}}
\caption{Distance $r$ in the equatorial plane in terms of the equatorial radius 
$R_{\rm eq}(\omega)$ at which recombination of He{\sc ii} (left panel) and 
H{\sc ii} (right panel) takes place for stars rotating with different 
velocities, indicated by $\omega$. The curves are (from top to bottom) for 
polar mass fluxes of: $F_{\rm m}({\rm pole}) = 
1.0\times 10^{-5}$, $1.5\times 10^{-5}$ and $2.0\times
10^{-5}$\,g\,s$^{-1}$cm$^{-2}$.}
\label{ion_he_h}
\end{figure}

\begin{figure}%[t!]
\resizebox{\hsize}{!}{\includegraphics{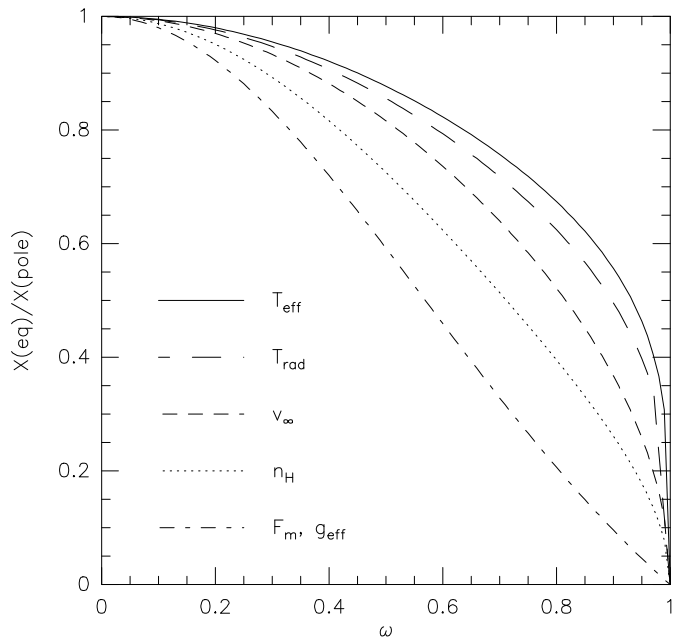}}
\caption{Surface parameters in the equatorial plane, normalized to their polar
values, as functions of the rotational velocity. $n_{\rm H}$ is the density at 
any distance.}
\label{eq}
\end{figure}

We calculated the equatorial recombination distance of H and He for stars with 
a large range in polar mass fluxes, $F_{\rm m, pole}$. For each mass flux we 
considered rotation velocities $\omega$ covering the complete range from 0 to 
1. In Fig.\,\ref{ion_he_h} we show the results of three representative models, 
calculated for $F_{\rm m, pole}=1.0\times 10^{-5}$, $1.5\times 10^{-5}$,
and $2.0\times 10^{-5}$g\,s$^{-1}$cm$^{-2}$. The recombination radii for
helium (left panel) and hydrogen (right panel) are given in units of the 
corresponding $\omega$--dependent stellar equatorial radius, $R_{\rm eq}
(\omega)$. Helium recombines for all models already close to the stellar 
surface. The recombination distance is found to decrease steadily with 
increasing stellar rotation, reaching the stellar surface for $\omega \ga 0.7$. 
This means that for rapidly rotating stars helium is neutral at the 
stellar surface. For hydrogen, the situation is different. For polar mass 
fluxes $F_{\rm m, pole} \la 1.5\times 10^{-5}$g\,s$^{-1}$cm$^{-2}$ the 
recombination distances decrease with increasing $\omega$, reach a minimum
in the range $\omega \simeq 0.70\ldots 0.75$, and increase again for higher
rotation speeds. If the input polar mass flux is higher than $1.5\times 
10^{-5}$g\,s$^{-1}$cm$^{-2}$, hydrogen shows the same trend as helium, i.e.
the recombination distances decrease steadily for increasing stellar rotation.
They reach also the surface of the star, but for rotation speeds $\omega \ga
0.85$. The models with $F_{\rm m, pole}=1.5\times 10^{-5}$g\,s$^{-1}$cm$^{-2}$
seem to be the case in ``transition". They show a kink at $\omega = 0.75$ and a 
subsequent steep drop in recombination distance. 

What causes the minimum and especially the strong increase in
hydrogen recombination distance for the lower mass flux models? 
This rather unexpected behaviour can be understood upon inspection of 
Fig.\,\ref{eq}. There we plotted the variation of the individual equatorial 
surface parameters with $\omega$, and of special interest are the radiation 
temperature and the particle density. Fig.\,\ref{eq} shows that with increasing 
rotation velocity, the density drops much
quicker from pole to equator than the temperature. For recombination to take
place right above the stellar surface, the number of ionizing photons has to be
reduced. This can be done either by decreasing the radiation temperature, or by
increasing the equatorial surface density and hence the optical depth. Since
the decrease in radiation temperature is determined by the rotation velocity
(with a fixed input polar value) we can only increase the input polar mass flux
to achieve a higher surface density for a given rotation velocity. A higher
density also has the effect of triggering recombination, this is however
only a secondary effect while the blocking of the raidation field is the
more important one. The density
in the top model shown in Fig.\,\ref{ion_he_h} is no longer high enough
for stars with $\omega \ge 0.7$ to absorb the ionizing photons
provided by the still rather high radiation temperature. Therefore, ionization
takes over again and shifts the recombination distance for higher rotation
velocities further out. Whether recombination takes place close to the star
therefore sensitively depends on the chosen input parameters of the rotating
star.

\section{Discussion}\label{discuss}

For our calculations we made a few assumptions, and we briely discuss 
their influence on the model results:

{\bf The electron temperature.}
The winds of hot stars are known to start with an electron temperature of about 
$0.8\,T_{\rm eff}$ at the stellar surface. Further out, they cool quickly 
(within a few stellar radii) and converge towards a more or less constant 
(terminal temperature) value (see e.g. Drew \cite{Drew}). The adoption of a 
constant (in radial direction) and maximum (i.e. $T_{\rm e}(\theta,r) = 
0.8\,T_{\rm eff}(\theta)$) electron temperature reduces (or suppresses) 
the total number of recombinations taking place, because the recombination 
coefficient is small for high temperatures but increases with decreasing 
temperature (see Fig.\,2 in Paper\,1). Our assumption therefore inhibits 
recombination and shifts the equilibrium of the ionization balance in favour of
the ionization of the wind material. This means that we have
overestimated the recombination distance.   

{\bf The wind velocity.} The velocity distribution in winds of hot stars can 
usually be approximated by a $\beta$-law (see e.g. Lamers \& Cassinelli 
\cite{LamersCassinelli}) which describes the increase in velocity from the 
small surface value to the terminal velocity. First, using the maximum (i.e. 
terminal) velocity instead of the more realistic $\beta$-law increasing 
velocity distribution results in an underestimation of the density, especially 
at distances close to the star. Second, the chosen value of 2000\,km\,s$^{-1}$
for the polar terminal velocity is rather high for a B-type supergiant and 
results equally in an underestimation of the density. Both assumptions 
therefore lead to an underestimation of the optical depth seen by the stellar 
radiation and an overestimation of the recombination distance.

{\bf The optical depth.} We calculate the optical depth properly only along
the $y$-axis. Since the surface density of a rotating star drops from pole
to equator and since the distance from the stellar surface to any point along 
the $y$-axis is the shortest one over which stellar photons can be absorbed,
this optical depth value is the smallest. Adopting this minimum optical depth 
for all directions towards the stellar surface therefore underestimates the real
optical depth and allows more ionizing photons to penetrate the wind material
to larger distances. This leads to an overestimation of the
recombination distance.

{\bf The OTS approximation.} In the OTS approximation it is assumed that every 
photon created via recombination and able to (re-)ionize hydrogen or helium
will be absorbed in the close vicinity of its generation location, and none 
will escape from the wind. These leads to additional 
ionizing photons (better known as the diffuse radiation field) at any location 
in the wind. The OTS approximation, even if not fully applicable
for the lower density regions, tends to overestimate the number of available
photons everywhere in the wind and therefore favours the ionization of the 
wind material. 

All these assumptions and simplifications made in our computations and listed
here show the tendency to overestimate the number of ionizing photons available 
at a certain location in the wind. Consequently, the recombination radius 
is shifted away from the star which means that our calculated distances at 
which the material is found to recombine are {\sl upper limits}.

\section{Applicability to the B[e] supergiants}\label{applic}

In Sect.\,\ref{results} we showed that the equatorial winds of rapidly rotating
stars might be neutral in hydrogen right from the stellar surface, even 
though the density in the equatorial wind is much lower than in the polar 
regions. Since our main goal is to find possible formation mechanisms
for hydrogen neutral disks around B[e] stars and especially B[e] supergiants,
we discuss here how reliable the results are and whether they are indeed
applicable to the known B[e] supergiants.

{\bf The rotation velocities of B[e] supergiants.} Our model is based
on the assumption that B[e] supergiants are rapidly rotating stars. What is
the evidence for their rapid rotation? In fact not much is known about their
rotation velocities. Due to their high density circumstellar medium, most of 
them do not show any photospheric absorption lines which might be used to 
derive a possible rotation speed. There are, however, three (out of 15) B[e] 
supergiants in the Magellanic Clouds for which photospheric absorption lines 
have been detected. From these line profiles only the projected stellar rotation
(i.e. $v \sin i$) can be derived with high accuracy. With their (often) poorly 
known inclinations $i$, only {\sl lower limits} to the real rotation speeds 
could be derived. These were found to be $\omega > 0.35$ and $\omega > 0.45$ 
for the two LMC B[e] supergiants \object{Hen\,S93} and \object{R\,66} 
(Gummersbach et al.\,\cite{Gummersbach}; Zickgraf \cite{Zickgraf2006}, 
respectively), and $\omega \simeq 0.8$ for the SMC B[e] supergiant 
\object{R\,50} (Zickgraf \cite{Zickgraf2000}). Especially the latter seems 
to rotate at a substantial amount of its critical velocity providing the 
basis for our research, although we cannot generalize that all B[e] supergiants 
are rapidly rotating.

{\bf Stellar luminosities and effective temperatures.}
A summary of the stellar parameters ($T_{\rm eff}$, $L_{*}$, $R_{*}$) of the MC 
B[e] supergiants is given e.g. in Zickgraf (\cite{Zickgraf2006}). From this list
it is obvious that the stellar luminosities and effective temperatures of the 
B[e] supergiants cover the range $10^{4} \la \log L/L_{\odot} \la 10^{6}$ and 
$10\,000 \la T_{\rm eff}[{\rm K}] \la 27\,000$, and our chosen values of the 
effective temperature fall well into this range while our luminosities are taken
as maximum values. The literature values for the B[e] supergiant effective 
temperatures, however, should be taken with caution because they
have mainly been derived from fitting Kurucz model atmospheres to the observed 
spectral energy distributions (see Zickgraf \cite{Zickgraf98} and references 
therein). However, these model atmospheres have been calculated under the 
assumption of spherically symmetric, non-rotating (i.e. uniformly bright) stars.
If B[e] supergiants are indeed rapidly rotating, then for a proper 
determination of the mean (i.e. observable) effective temperature 
a comparison with composite spectra should be undertaken (see e.g. Lovekin et 
al.\,\cite{Lovekin}). Of course, to do so the inclination and the rotational 
velocities must be known which is usually not the case for B[e] supergiants.

{\bf Mass loss rates.} The models presented in Fig.\,\ref{ion_he_h} are for 
stars with polar mass fluxes between $1.0\times 10^{-5}$ and $2.0\times 
10^{-5}$\,g\,s$^{-1}$cm$^{-2}$. Therefore, the range in mass loss rates 
covered by these calculations extends from $\dot{M}_{\rm min} = 3.4\times 
10^{-5}$M$_{\odot}$yr$^{-1}$ (critically rotating star with lower polar 
mass flux) to $\dot{M}_{\rm max} = 1.3\times 10^{-4}$M$_{\odot}$yr$^{-1}$ 
(non-rotating star with higher polar mass flux), and is in good agreement 
with the known mass loss rates for MC B[e] supergiants, which range from 
about $10^{-5}$M$_{\odot}$yr$^{-1}$ to about $10^{-4}$M$_{\odot}$yr$^{-1}$
(see Zickgraf \cite{Zickgraf2006}).

{\bf The equatorial surface density.} 
According to Eq.\,(\ref{masscont}), the surface density of a non-rotating star
behaves as $n_{\rm H}(R_*) \sim \dot{M}/(R_{*}^{2}v(R_*)) = F_{\rm m}/v(R_*)$.
For our model calculations we have chosen mass loss rates in the range reliable
for B[e] supergiants, but we used some maximum values for the velocity ($v(R_*) 
= v_{\infty}$) and stellar radius\footnote{There exists one exception in the 
literature: The (more or less) pole-on LMC star \object{R\,66} is assumed 
to have a radius of 125\,R$_{\odot}$ (see Table\,1 in Zickgraf 
\cite{Zickgraf2006}), while all other stars fall well below our adopted radius 
of 82\,R$_{\odot}$.}. Therefore, for non-rotating stars the surface densities 
in our calculations provide some lower limits. 

For (especially rapidly) rotating stars, the situation becomes more complicated.
Now, the surface density is found to drop from pole to equator. On the other
hand, B[e] supergiants are supposed to have circumstellar disks. These disks
are much denser than what is found for the polar wind regions, with density
contrasts on the order of $\rho_{\rm eq}/\rho_{\rm pole}\simeq 100\ldots 1000$.
Models proposed to explain the formation of these high density disks are the
bi-stability mechanism (Lamers \& Pauldrach \cite{LP}; Pelupessy 
\cite{Pelupessy}) and wind compression (Bjorkman \& Cassinelli \cite{BC}). 
While the bi-stability mechanism in a rotating star might account
for an increase by a factor of $\sim 10$ in equatorial density only, the wind
compression, especially for rapidly rotating stars (or more precisely the flow
of material towards the equatorial plane), can be inhibited due to the 
appearance of a non-radial force provided by the radiation (Owocki et al. 
\cite{OCG}). Recently, the existence of a slow solution
in line--driven winds of rapidly rotating stars has been found (Cur\'{e} 
\cite{Cure04}; Cur\'{e} et al. \cite{Cure05}). Inclusion of the bi-stability
jump resulted in an equatorial density enhancement (at least in the 
close vicinity of the star) by a factor of 100 -- 1000, just what is needed
to explain the disks of B[e] supergiants. However, these solutions have been
found adopting a spherically symmetric star and neglecting gravity darkening,
and it still needs to be confirmed that these slow solutions
will also exist when gravity darkening is taken properly into account.
 
Our models do not account for any density enhancements either due to 
bi-stability or due to wind compression. Such an increase in equatorial 
surface density by a factor 100 -- 1000 (for the same imput values) would 
result in a recombination distance even closer to the stellar surface, or 
would mean that the polar mass flux of the model star can be reduced by the 
same factor and the material would still recombine in the vicinity of the star. 
Such a lower mass flux (and hence a lower mass loss rate) might be desirable 
if the winds of B[e] supergiants are clumped. A clumpy wind in contrast to the
assumed smooth density distribution is found to overestimate the mass loss 
rates (derived e.g. from H$\alpha$) by a factor of 10 or more 
(see e.g. Hillier \cite{Hillier05}; Bouret et al. \cite{Bouret}). 

Even if the winds of B[e] supergiants will turn out to have lower mass loss 
rates, the ionization structure calculations presented in this paper show  
that recombination of the equatorial wind material of rapidly rotating stars
can take place at or at least close to the stellar surface.

%%%%%%%%%%%%%%%%%%%%%%%%%%%%%%%%%%%%%%%%%%%%%%%%%%%%%%%%%%%%%%%%%%%%%%%%%

\section{Conclusion}\label{conclusions}

We investigated the influence of rigid rotation on the surface and wind
parameters of hot luminous stars, with emphasis on the non-spherical winds 
of B[e] supergiants. Since B[e] supergiants are known to have equatorial disks 
that show evidence for hydrogen neutral material in the 
vicinity of the stellar surface, the calculations are restricted to the 
equatorial plane. Due to the symmetric stellar radiation field (with respect 
to the equatorial plane) the problem of finding the recombination distance 
reduces even to the 1-dimensional case. The radiation field is however treated
2-dimensionally, to properly account for the latitude dependences of the 
parameters like effective temperature, wind velocity and density.
The ionization balance equations are solved in a pure hydrogen plus helium wind.
All assumptions made during our calculations have in common that the number
of available ionizing photons at any location within the equatorial plane is
overestimated. This results in a shift of the recombination distance
to larger values, which means that we have calculated {\sl upper limits} for 
the recombination distance.

The major result is that despite the drop in equatorial surface density with
increasing rotation velocity (neglecting any possible equatorial density
enhancement due to bi-stability and/or wind compression), hydrogen recombines 
at (or close to) the stellar surface for rotating models with a polar mass flux
$F_{\rm m, pole}\ga 1.5\times 10^{-5}$g\,s$^{-1}$cm$^{-2}$ and rotation
velocities $\omega \ga 0.8$ (see Fig.\,\ref{ion_he_h}).
These mass fluxes correspond to mass loss rates $\dot{M} \ga 5\times 
10^{-5}$\,M$_{\odot}$yr$^{-1}$ for our chosen model stars with supergiant 
stellar and wind parameters. Since the mass loss rates for B[e] supergiants are
found to lie in the range $\dot{M}=10^{-5}\ldots 10^{-4}$\,M$_{\odot}$yr$^{-1}$
we can expect that at least some of these stars might have hydrogen 
neutral equatorial material close to their stellar surface, given that they
are indeed (rapidly) rotating stars, as is found for at least three of them.

%%%%%%%%%%%%%%%%%%%%%%%%%%%%%%%%%%%%%%%%%%%%%%%%%%%%%%%%%%%%%%%%%%%%%%%%%

\begin{acknowledgements}

%We thank the unknown referee for critical comments that have led 
%to an improvement of this paper. 
I am grateful to Henny Lamers for his helpful comments and suggestions and 
for his proof-reading of  
the draft of this paper. This research was supported by grants from GA \v{C}R 
205/04/1267 and from the Nederlandse Organisatie voor Wetenschappelijk 
Onderzoek (NWO) grant No.\,614.000.310.

\end{acknowledgements}

%%%%%%%%%%%%%%%%%%%%%%%%%%%%%%%%%%%%%%%%%%%%%%%%%%%%%%%%%%%%%%%%%%%%%%%%%

\end{document}